\documentclass[conference]{IEEEtran}
\IEEEoverridecommandlockouts
\usepackage{cite}
\usepackage{amsmath,amssymb,amsfonts}
\usepackage{algorithmic}
\usepackage{graphicx}
\usepackage{textcomp}
\usepackage[table]{xcolor}
\def\BibTeX{{\rm B\kern-.05em{\sc i\kern-.025em b}\kern-.08em
    T\kern-.1667em\lower.7ex\hbox{E}\kern-.125emX}}

\usepackage{graphicx}

\usepackage{pbox}
\usepackage{comment}
\usepackage{csvsimple}
\usepackage{fancyhdr}

\usepackage{soul}
\usepackage{mathtools}
\usepackage[normalem]{ulem}
\usepackage{url}
\usepackage{hyperref}
\usepackage{lscape}
\usepackage{numprint}
\usepackage{array}
\usepackage{multicol}
\usepackage[caption=false]{subfig}
\usepackage{float}
\usepackage{subfloat}
\usepackage{mathtools}
\usepackage{caption}
\usepackage{amsmath}
\usepackage{makecell}
\usepackage{tikz}

\usepackage{siunitx}
\usepackage{pgfplotstable}
\newcommand\highlight[1]{\textcolor{black}{#1}}
\newcommand\iiswc[1]{\textcolor{black}{#1}}

\renewcommand\fbox{\fcolorbox{blue}{white}}

\begin{document}

\title{Performance Analysis of Scientific Computing Workloads on Trusted Execution Environments}


\author{\IEEEauthorblockN{Ayaz Akram}
\IEEEauthorblockA{
\textit{UC Davis} \\
yazakram@ucdavis.edu}
\and
\IEEEauthorblockN{Anna Giannakou}
\IEEEauthorblockA{
\textit{LBNL} \\
agiannakou@lbl.gov}
\and
\IEEEauthorblockN{Venkatesh Akella}
\IEEEauthorblockA{
\textit{UC Davis} \\
akella@ucdavis.edu}
\and
\IEEEauthorblockN{Jason Lowe-Power}
\IEEEauthorblockA{
\textit{UC Davis} \\
jlowepower@ucdavis.edu}
\and
\IEEEauthorblockN{Sean Peisert}
\textit{LBNL \& UC Davis} \\
sppeisert@lbl.gov}

\maketitle

\begin{abstract}
Scientific computing sometimes involves computation on sensitive data.
Depending on the data and the execution environment, the HPC (high-performance computing) user or data provider may require confidentiality and/or integrity guarantees.
To study the applicability of hardware-based trusted execution environments (TEEs) to enable secure scientific computing, we deeply analyze the performance impact of AMD SEV and Intel SGX for diverse HPC benchmarks including traditional scientific computing, machine learning, graph analytics, and emerging scientific computing workloads.
We observe three main findings: 1) SEV requires careful memory placement on large scale NUMA machines (1$\times$--3.4$\times$ slowdown without and 1$\times$--1.15$\times$ slowdown with NUMA aware placement), 2) virtualization---a prerequisite for SEV---results in performance degradation for workloads with irregular memory accesses and large working sets (1$\times$--4$\times$ slowdown compared to native execution for graph applications) and 3) SGX is inappropriate for HPC given its limited secure memory size
and inflexible programming model (1.2$\times$--126$\times$ slowdown over unsecure execution).
Finally, we discuss forthcoming new TEE designs and their potential impact on scientific computing.
\end{abstract}


\section{Introduction and Background}\label{sec:introduction}
High performance computing (HPC) operators provide the computing hardware, storage, and networking infrastructure to help researchers solve large-scale computational problems.
Increasingly, scientific computing involves data analysis, rather than traditional modeling and simulation.
The data being analyzed is often provided by third parties to individual researchers or the HPC center for use by multiple users.
The data may be fully open, or may be in some way sensitive e.g., subject to government regulations, considered proprietary, or pertain to public privacy concerns.
This paper analyzes the performance impact of protecting this data with hardware-based trusted execution environment (TEE) on AMD and Intel based systems, one aspect of a larger secure system design~\cite{peisert2017security}.

The use of sensitive data at HPC centers raises security concerns from both external threats (the user of compute resources and the data provider might not trust other users sharing the compute resources) and internal threats (the user and the data provider might not trust the compute provider).
Data owners (or compute providers) have to make a tradeoff between accepting the risk of hosting (or providing compute for) the sensitive data and declining the data.
Currently, where secure HPC enclaves exist, such as those currently being used for analysis of sensitive health-related data pertaining to COVID-19 at universities and research institutes around the world, trust required by data providers of the compute facility, and liability to an organization for hosting such data are both very high.
Moreover, traditionally, the environments created to protect sensitive data have significant usability challenges.
For example, processing sensitive health data requires dramatically different environments compared to those typically used in NSF high-performance computing facilities.
Processing capabilities are limited to only a handful of racks and access requires virtual private networks (VPNs) and/or remote desktops.
These onerous usability requirements are particularly cumbersome for the scientific community which is mostly used to working in very open, collaborative, and distributed environments.

\begin{figure}
  \centering
  \includegraphics[width=0.95\linewidth]{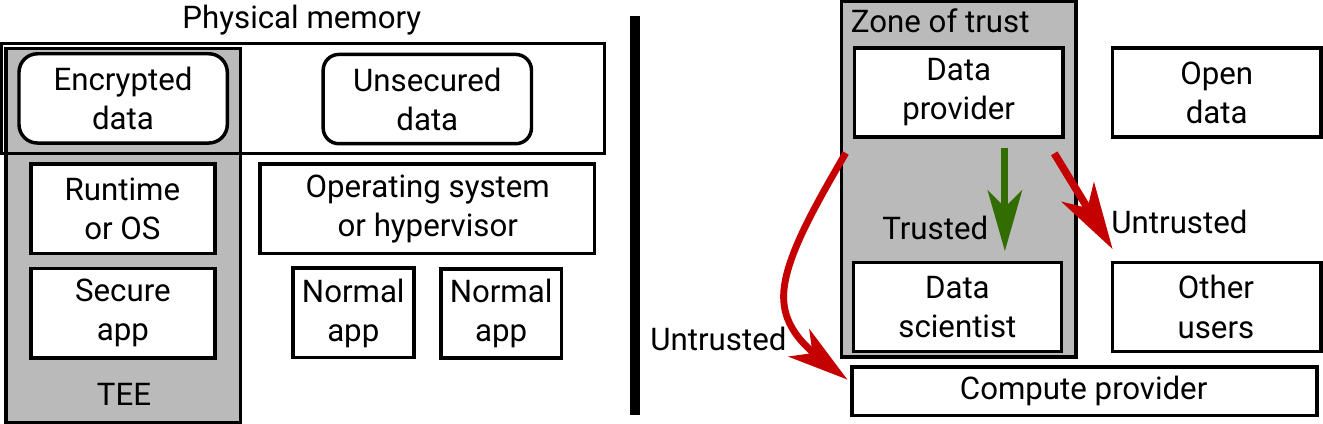}
  \caption{Trusted execution environments and threat model}
  \label{fig:tees}
\end{figure}

The use of computing hardware supported by trusted execution environments (TEEs), which provide confidentiality and/or integrity guarantees for the sensitive data and compute, can act as a first step to enable secure scientific computing without severely impacting the usability of HPC systems by providing equivalent or greater layers of protection with fewer steps for the end-user.
And, data owners (or computer providers) are not required to make any tradeoffs between increased risk and declining to host data in a TEE based secure computing environment.

Figure~\ref{fig:tees} shows how TEEs fit into the system architecture.
TEEs can be used to isolate data from other users and the operating system or hypervisor that is controlled by the compute provider.
Examples include Intel's SGX (Software Guard Extension)~\cite{costan2016intel}, ARM's TrustZone~\cite{aTZ}, AMD's SEV (Secure Encrypted Virtualization)~\cite{kaplan2016amd}, and emerging research platforms like RISC-V's Keystone~\cite{2019arXiv190710119L}.
\iiswc{In this paper, we investigate the performance of traditional and emerging HPC applications on TEEs based computing hardware.}
Researchers have evaluated the performance of TEEs in the context of client and cloud computing~\cite{weisse2017regaining,gjerdrum2017performance}.
However, to the best of our knowledge, a detailed analysis of performance overhead of TEEs on realistic HPC workloads has not been conducted in the past.
The goal of this paper is to fill that gap.
More specifically, we evaluate the performance of AMD's SEV and Intel's SGX platforms for a range of different HPC applications including traditional scientific computing (via the NAS Parallel Benchmarks), graph processing (via the GAP Benchmark Suite~\cite{beamer2015gap}), and emerging potentially sensitive workloads, genetic analysis (BLAST~\cite{altschul1990basic}), traffic simulation (Mobiliti~\cite{chan2018mobiliti}), \iiswc{hydrodynamics	stencil	calculation (LULESH~\cite{LULESH:spec}), and a particle transport simulation (Kripke~\cite{kunen2015kripke})} on realistic datasets.

The main contributions of this work are in the form of three major findings:

1) \textit{SEV can be used for secure scientific computing without significant performance degradation for most workloads if it is configured correctly.}
We found that when using the default configuration of SEV, applications suffer some performance degradation (up to 3.4$\times$) from poor NUMA data placement as all data resides on a single NUMA node.
However, most of this performance loss can be mitigated by simply configuring the system to use NUMA interleaving.

2) \textit{Irregular workloads performance suffers due to virtualization when running under SEV.}
SEV requires running applications in a virtualized environment (e.g., QEMU), which causes performance loss for highly-irregular workloads such as graph analytics and workloads with many I/O operations.
Future architectures and systems should concentrate on reducing this overhead to improve the performance of HPC applications in secure environments.

3) \textit{SGX is inappropriate for scientific computing.} We found that running \iiswc{unmodified HPC} applications under SGX results in a significant performance degradation.
We observed slowdowns from 1.2$\times$ to 126$\times$ compared to unsecure execution. This slowdown is mainly due to the limited secure memory capacity (about 100~MB) and the fact that HPC applications have large working sets and are difficult to partition into secure and normal components. \iiswc{Secondly, the HPC workloads under SGX exhibit poor multithreaded scaling (an important characteristic of HPC workloads).} Furthermore, we found that the support for application development with SGX is also limited, which makes it inadequate (or at least unsuitable) for HPC environments, where domain scientists often rely on unmodified off-the-shelf third-party libraries such as Tensorflow.

\section{Threat model} \label{threatmodel}

We assume that HPC system administrators are not trusted and that host operating systems and hypervisors are not trusted.
However, the guest operating system of a virtual machine which is owned by the user is trusted.
We assume very simple physical attacks are within scope, but that physical attacks that are more time consuming, such as opening a rack-mount HPC system and removing chips soldered on the board, are less important at this time because there are other means, such as video cameras pointed at the HPC systems, to monitor and mitigate such attacks.
We assume HPC users themselves are trusted to not exfiltrate their own data, \iiswc{though we do not trust them to not attack others.}
Also, we focus on general-purpose computing hardware---FPGAs, GPUs, dedicated ASICs are not considered in this paper, mainly because no commercial TEEs yet exist for these hardware accelerators.

We assume that data providers trust the data users or that some other means (e.g., differential privacy~\cite{dwork2006differential}) will ensure the sensitive data is not improperly exfiltrated by the scientific application developers and users.
Figure~\ref{fig:tees} shows how TEEs fit into this threat model.

\section{Background and Prior Work} \label{background}

\subsection{Trusted Execution Environments}


Trusted Execution Environments in hardware, at minimum, provide some degree of hardware-enforced separation from other users and processes, and the ability of end users to verify through cryptographic attestation that execution is taking place within the TEE.
Some TEEs, including Intel's Software Guard Extensions (SGX) and AMD's Secure Encrypted Virtualization (SEV), also support encrypted memory.
Both SGX and SEV protect against malicious system administrators and host operating systems.
TEEs have their roots in earlier cryptographic hardware functions, including Trusted Platform Modules.
In this work, we analyze the performance of AMD SEV~\cite{kaplan2016amd} and Intel SGX~\cite{costan2016intel}.
We exclude the other major commercially available option ARM TrustZone~\cite{aTZ} from this study as existing TrustZone based TEEs mainly target embedded and mobile devices, not general purpose compute devices~\cite{sandro2019trustzone}.
Table \ref{tab:feature-comparison} shows a short feature comparison of both SGX and SEV.

\subsubsection{Intel Software Guard Extensions (SGX)}
\label{sec:sgxback}


Intel SGX divides the application into two code segments, untrusted and trusted (enclave) which cannot directly communicate and interact.
Only the trusted part is allowed to access confidential data residing in encrypted form in a memory region called Enclave Page Cache (EPC).
The need to split an application (manually) into trusted and untrusted parts can be a challenging task for HPC applications as they often rely on many third-party libraries.
The size of the EPC is set to be 128MB, out of which almost 32MB is used to store the metadata needed to provide security guarantees~\cite{taassori2018vault}.
In case of SGX, the MEE (memory encryption engine) which sits besides the memory controller on the CPU package is responsible for permission checks for EPC accesses, provision of data confidentiality by encrypting the data when it leaves the CPU package to reside EPC and performs integrity tree operations on the data stored in the EPC.

Both parts of an SGX application communicate through an interface of in/out calls (\textit{ecall/ocall}).
\textit{ecall} and \textit{ocall} perform a secure context switch which includes: enabling/disabling of tracing mechanisms, permission checks for enclave memory, validation of enclave control structures and backing up/reloading of registers that represent untrusted execution context \cite{weisse2017regaining}.
Similarly, enclave code cannot use normal system calls directly, rather the control needs to be transferred to the non-secure part of the application first using \textit{ocall}.
SGX requires application changes and/or recompilation.
However, there are third-party solutions (e.g. SCONE~\cite{arnautov2016scone}), which allow running unmodified workloads, but they have their own limitations (discussed in section~\ref{sec:sgxperf}).
SGX also provides integrity guarantees through the use of integrity trees consisting of counters to keep track of version of EPC pages to protect against replay attacks.

\subsubsection{AMD Secure Encrypted Memory (SEV)}
\label{sec:sevback}

In case of SEV, the protected memory can be equal to the size of the entire physical memory.
AMD SEV provides transparent encryption of memory used by virtual machines (unique encryption key associated with each isolated guest).
As a result, SEV has a larger trusted computing base (TCB), compared to SGX, which includes \highlight {the guest} OS, the hypervisor, and the CPU package.
In contrast to SGX, which requires application modifications, SEV does not require changes in an application's code.
However, the application needs to be run inside a VM managed by the hypervisor (QEMU).
SEV lacks integrity support and does not provide protection against replay attacks.

\begin{table}
  \centering
     \caption{Feature Comparison}
    \begin{tabular}{|l|l|l|}
    \Xhline{2\arrayrulewidth}
     \textbf{Feature} & \textbf{SGX} & \textbf{SEV} \\ \Xhline{2\arrayrulewidth}
    Integrity Provision & \cellcolor{green!25} Yes & \cellcolor{red!25} No  \\ \hline
    TCB Size & \cellcolor{green!25} Small & \cellcolor{red!25} Large  \\ \hline
    Secure Memory Size & \cellcolor{red!25} 128 MB & \cellcolor{green!25} Up to RAM size  \\ \hline
    Application Changes & \cellcolor{red!25} Required & \cellcolor{green!25} Not Required \\ \Xhline{2\arrayrulewidth}

  \end{tabular}
  \label{tab:feature-comparison}
  \end{table}

\begin{table}
  \centering
     \caption{System Configurations. See Figure~\ref{fig:numa} for details on the two EPYC systems.}
    \begin{tabular}{|p{0.75cm}|p{1.55cm}|p{1.5cm}|p{1.55cm}|p{1.5cm}|}
    \Xhline{2\arrayrulewidth}
    \textbf{Feature} & \textbf{AMD SEV 1} & \textbf{AMD SEV 2} & \textbf{AMD SEV 3} & \textbf{Intel SGX} \\ \Xhline{2\arrayrulewidth}
    \textbf{CPU} & \footnotesize{EPYC 7401P} & \footnotesize{EPYC 7702} & \footnotesize{EPYC 7402P} & \footnotesize{Core i7-8700} \\ \hline
    \textbf{Sockets} & 1 & 2 & 1 & 1 \\ \hline
    \textbf{Cores} & 24 & 128 & 24 & 6 \\ \hline
    \textbf{NUMA} & 4 Nodes & 2 Nodes & 1 Node & 1 Node \\ \hline
    \textbf{RAM} & 64GB & 1TB & 64GB & 32GB  \\ \Xhline{2\arrayrulewidth}
  \end{tabular}
  \label{tab:Systems}
  \end{table}

\subsection{Prior studies of TEE performance}

There exist many prior studies on the performance impact of TEEs which either focus on a single TEE or a comparison of different TEEs~\cite{Mofrad2018-hm, akram2019using, dinh2019everything, gottel2018security, Brenner2018-ek, gottel2018security}.
These studies found that while the SGX overhead can be significant it can be mitigated for \emph{small or partition-able workloads} and that the SEV overhead is negligible for small workloads.
However, these studies are different from our work in three fundamental ways: (1) none of these works focus on scientific/high-performance computing, (2) they target micro-benchmarks or small benchmarks which are usually single-threaded, (3) with reference to SEV, none of these works evaluate large machines with multiple NUMA nodes which are common in HPC environments.

Brenner et al. evaluated cloud applications under SGX and found that due to the limited enclave memory capacity, performance was best when using many enclaves in a distributed-computing pattern~\cite{Brenner2018-ek}.
Genie~\cite{Zhangy2018-sq} and Slalom~\cite{tramer2018slalom} evaluate AI inference applications in TEEs.
Both of these works make significant changes to the application code to work with SGX, unlike this work which focuses on evaluating a broad range of unmodified HPC applications.
Partitioning of applications into secure and non-secure parts is a difficult task, especially in HPC settings as HPC workloads often rely on various third-party libraries.
There has also been prior work (e.g., Glamdring~\cite{Lind2017-un}) which is a framework for automatic application partitioning into secure and non-secure parts.
We did not evaluate Glamdring for HPC applications since for most of these applications \emph{all of their data} would be classified and confidential and there is little to partition.

There has also been prior work focused on improving the software and hardware architecture of SGX to improve performance~\cite{taassori2018vault, Saileshwar2018-pu, Orenbach2017-ks, gjerdrum2017performance}.
We do not evaluate these research proposals in this paper.
Instead, we focus on the currently available hardware technologies.

\subsection{Beyond TEEs}

Fully homomorphic encryption~\cite{gentry2009fully}, secure multi-party computation~\cite{yao1986generate}, and functional encryption all represent methods for computing over encrypted data by leveraging software algorithms, rather than hardware properties.
Similar protection properties would apply in these cases, but with two important caveats and one potential benefit: first, these techniques are computationally expense and are therefore significantly slower than hardware TEEs.
This is true even though performance has improved from being on the order of 1 trillion times slower than computing in cleartext ten years ago to perhaps only ten to a hundred times slower than computing in cleartext, depending on the technique used and the operations needing to be computed under encryption.
For example database searches have been shown to be relatively fast~\cite{popa2012cryptdb,poddar2016arx,FullerVYSHGSMC17,titus2018sig}, but operations requiring both addition and multiplication are much slower.
The second caveat is that programs typically need very significant modification to use this technique, often causing each application of the technique to require extensive adaptation of the underlying cryptographic approach.
A potential benefit is that leveraging some of these approaches could allow the threat model to be expanded to include protecting malicious users.

That said, TEEs could also be used to protect against malicious users by incorporating differential privacy.
Indeed, a ``complete'' architecture that we envision is one in which ``sensitive'' data cannot be computed upon unless inside the TEE, and similarly, that sensitive data cannot be output unless via the TEE, which also enables output to be forced to protected through some kind of ``guard'' or gating policy, such as differential privacy~\cite{dwork2006differential}.
\section{Methodology}
\label{methodology}

We picked traditional scientific computing workloads as well as modern applications which fit the criteria of HPC application domain.
Table~\ref{tab:workloads} provides a summary of the workloads evaluated in this work.

\subsection{Traditional HPC Benchmarks/Kernels (NPB)}

We evaluate workloads traditionally used to benchmark HPC systems i.e. NAS Parallel Benchmark suite (NPB)~\cite{bailey1991parallel}.
The NAS Parallel Benchmark suite, consisting of different kernels and pseudo applications, has been used to study HPC systems for a long time and is still getting updates.
These benchmarks can be used with multiple input data sizes, thus different class names.
In this work, we used NPB Class C for both SEV and SGX and NPB Class D for SEV only.

\begin{table}
  \centering
     \caption{Details of workloads evaluated.}
    \begin{tabular}{|p{1.5cm}|p{3.8cm}|p{2.1cm}|}
    \Xhline{2\arrayrulewidth}
    \multicolumn{3}{|c|}{ \textbf{NAS Parallel Benchmarks NPB}} \\
    \Xhline{2\arrayrulewidth}
     \textbf{Benchmark} &   \textbf{Description} &  \textbf{Working-Set \highlight{(Class C \& D)}} \\
    \Xhline{2\arrayrulewidth}
     bt &  block tri diagonal solver &  0.68 \& 10.67 GB\\ \hline
     cg &  conjugate gradient &  0.36 \& 16.31 GB \\ \hline
     ep &  embarrassingly parallel &  0.028 \& 0.028 GB \\ \hline
     is &  integer sorting &   1.03 \& 33.1 GB \\ \hline
     lu &  lower-upper gauss-seidel solver &  0.59 \& 8.89 GB \\ \hline
     mg &  multi-grid method &  3.3 \& 26.46 GB \\ \hline
     sp &  scalar penta diagonal solver &  0.78 \&  11.62 GB \\ \hline
     ua &  unstructured adaptive mesh &  0.47 \& 7.30 GB \\ \Xhline{2\arrayrulewidth}

    \Xhline{2\arrayrulewidth}
    \multicolumn{3}{|c|}{ \textbf{GAP Benchmark Suite~\cite{beamer2015gap}} (road network)} \\
    \Xhline{2\arrayrulewidth}
     \textbf{Benchmark} &   \textbf{Description} &  \textbf{Working-Set} \\
    \Xhline{2\arrayrulewidth}
     bc &  betweenness centrality &  1.15 GB\\ \hline
     bfs &  breadth first search &   0.97 GB \\ \hline
     pr &   page rank &  0.97 GB \\ \hline
     sssp &  single-source shortest paths &   1.39 GB \\ \hline
     cc &  connected components &  0.96 GB \\ \hline
     tc &  triangle counting &  0.57 GB \\ \Xhline{2\arrayrulewidth}

    \Xhline{2\arrayrulewidth}
    \multicolumn{3}{|c|}{ \textbf{Other Modern HPC Workloads}} \\
    \Xhline{2\arrayrulewidth}
     \textbf{Benchmark} &   \textbf{Description} &  \textbf{Working-Set} \\
    \Xhline{2\arrayrulewidth}
     \iiswc{Kripke~\cite{kunen2015kripke}} &  \iiswc{Hydrodynamics	Stencil	Calculation} &  \iiswc{7.4 GB} \\ \hline
     \iiswc{LULESH~\cite{LULESH:spec}} &  \iiswc{Particle Transport Simulation} &  \iiswc{0.108 GB} \\ \hline
     LightGBM~\cite{ke2017lightgbm} &  Gradient Boosted Decision Tree Framework from Microsoft &  5.4 GB \\ \hline
     Mobiliti~\cite{chan2018mobiliti} &  Transportation System Simulator &  1.06 GB \\ \hline
     BLASTN~\cite{altschul1990basic} &  Basic Local Alignment Search Tool &  26.20 GB \\ \Xhline{2\arrayrulewidth}

  \end{tabular}
  \label{tab:workloads}
  \end{table}

\subsection{Modern and Emerging HPC Workloads}

Apart from the traditional scientific computing kernels/workloads, we also focus on workloads which characterize modern HPC usage.
We selected a set of graph workloads (GAPBS)~\cite{beamer2015gap} with an input of a graph of road networks in the US.
As a proxy for general machine learning training we used a decision tree workload (LightGBM)~\cite{ke2017lightgbm} (characterized by irregular memory accesses) which is trained using Microsoft's Learning to Rank (MSLR) data set.
~Finally, we used modern HPC workloads as well, including \iiswc{Kripke~\cite{kunen2015kripke} (a particle transport simulation), LULESH~\cite{LULESH:spec} (a hydrodynamics simulation)}, Mobiliti~\cite{chan2018mobiliti} (a transportation benchmark), and BLAST \cite{altschul1990basic} (a genomics workload).
\iiswc{Kripke~\cite{kunen2015kripke} is a highly scalable code which acts as a proxy for 3D Sn (functional discrete-ordinates) particle transport.
Livermore Unstructured Lagrange Explicit Shock Hydro (LULESH)~\cite{LULESH:spec} application solves a simple yet ``full-featured" hydrodynamics simulation problem.}
Mobiliti~\cite{chan2018mobiliti} is a transportation system simulator (based on parallel discrete event simulation), designed to work
on high performance computing systems.
Basic Local Alignment Search Tool (BLAST)~\cite{altschul1990basic} is a famous bioinformatics tool, which is used to search sequence similarity of a given genome sequence compared to an existing database.
We specifically use BLASTN in this work, which is a version of BLAST used to search a nucleotide sequence against a nucleotide database.

\subsection{Hardware Platforms Used}

\begin{figure}
  \subfloat[\scriptsize{AMD EPYC 7401P (\tiny{Naples})}]{
    \includegraphics[width=.32\linewidth]{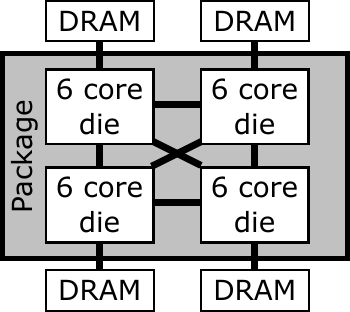}
    \label{fig:naples}
  }
   \subfloat[\scriptsize{AMD EPYC 7702 (\tiny{Rome})}]{
    \includegraphics[width=.64\linewidth]{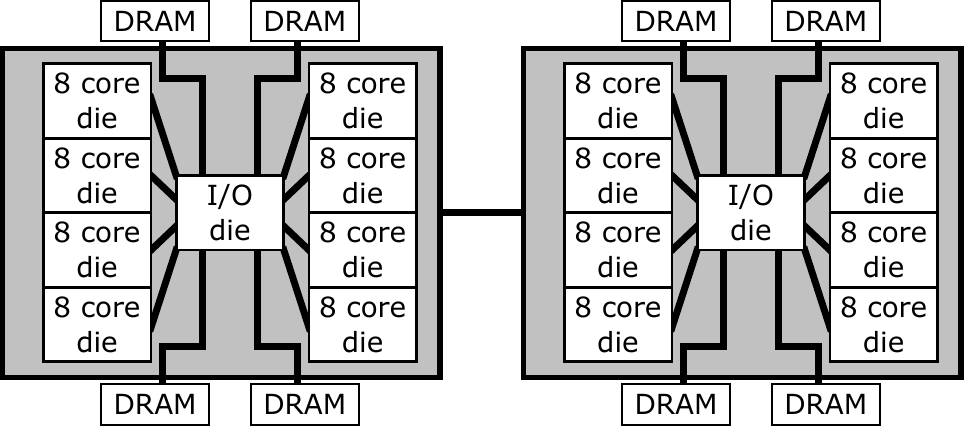}
    \label{fig:rome}
  }
  \caption{Details of the non-uniform memory architecture for the two AMD systems evaluated in this paper.}
  \label{fig:numa}
\end{figure}

Table~\ref{tab:Systems} shows the configurations of the hardware platforms used for these experiments.
For all of our evaluations, we evaluate \emph{without} hyperthreading by limiting the number of threads to the number of cores on each platform.

We used three server class AMD machines.
Figure~\ref{fig:numa} shows the detailed NUMA configuration of the AMD EPYC 7401P (Naples architecture, Figure~\ref{fig:naples}) and the AMD EPYC 7702 (Rome architecture, Figure~\ref{fig:rome}).
The Naples-based system has 24 CPU cores with 6 cores on each of four dies in a single multi-chip module.
Although this system is a single socket platform, it has four NUMA nodes.
A multi-chip module package has characteristics similar to a multi-socket system in terms of latency and bandwidth between separate dies.
With its four NUMA nodes the Naples-based system has high variation in memory latency depending on if the data is in the local NUMA node or one of the remote NUMA nodes.

We also evaluated a recent Rome-based system since this design has a more uniform memory architecture.
The Rome-based system has 64 cores with 8 cores on each of 8 dies in a multi-chip package, and it is a dual socket system for a total of 128 cores.
The Rome system has more chips per package, but has a \highlight{more uniform} memory architecture since each die is equidistant from the I/O die with the memory controllers.
In the Rome-based system we evaluated, there is only one NUMA node \emph{per socket}.
However, we used a dual socket system so our evaluations have two NUMA nodes.
We also used an EPYC 7402P (Rome architecture, with one socket) system for validation of some results discussed in section~\ref{sec:findings}.

The forthcoming supercomputers Frontier and El Capitan are based on AMD microarchitecture~\cite{AMD-Exascale}, though these will likely be based on a future microarchitecture (e.g., Zen 3 or Zen 4).
The specific memory architecture of these future devices are currently unknown publicly, but it is likely that they will support multiple sockets and thus will have at least as much non-uniformity as the Rome-based system.
Google's confidential cloud computing initiative also relies on AMD SEV for trusted execution support~\cite{google-conf}.

We use a desktop-class processor with 6 cores and a single NUMA node to perform Intel SGX experiments, as there did not exist a server-class Intel processor with the support of SGX at the time of performing SGX experiments in this paper.
Recently, Intel SGX is made available in one of the Intel Xeon parts (Xeon E3).
However, the size of secure memory (doubled to be 256MB in total) is still significantly smaller than the working set of most of the workloads studied in this paper (only \textit{ep} has a working set smaller than 256MB) and the conclusions drawn in this work (discussed in section~\ref{sec:findings}) should still hold true.

\subsection{Software Tools/Frameworks}

To execute unmodified applications under SGX, we make use of SCONE~\cite{arnautov2016scone} framework container.
Programs are compiled statically and linked against a modified standard C library in SCONE.
SCONE runtime also makes use of threads outside the enclave to perform asynchronous execution of system calls.
We evaluated other SGX interfaces and picked SCONE as it provided the most complete support for unmodified applications.
These other SGX programming interfaces are discussed in section~\ref{sec:findings} (Finding 4.4).
\iiswc{Rewriting HPC applications for SGX's programming model, by partitioning them into secure and un-secure components, is arduous but not impossible.
However, in this work we focus on the use case of unmodified HPC applications.}
\highlight{Also, the overhead of containerization like SCONE has been shown to be low.
The original work~\cite{arnautov2016scone} introducing SCONE showed that it has a 0.6--1.22$\times$ throughput compared to native execution for services like Apache, Redis, NGINX, and Memcached~\cite{arnautov2016scone}.
We also tested the performance of NAS parallel benchmarks in the ``simulation mode'' of SCONE.
This mode uses all of the SCONE interfaces, but does not enable SGX.
We found that the geometric mean of slowdown compared to native execution is 1.19$\times$, which is insignificant compared to the slowdown of trusted execution (with SGX) in SCONE as shown in section~\ref{sec:findings} (Finding 4).
}
Finally, we observed the performance of two memory intensive micro-benchmarks, partitioned into secure and un-secure parts directly using Intel SGX SDK, and found those numbers to be in line with our observations with SCONE as discussed in section~\ref{sec:findings} (Finding 4).

For SEV, we make use of the AMD provided scripts to set-up the SEV enabled host machine and the guest virtual machine (VM) managed by QEMU.
We also evaluated using Kata~\cite{kata} which is a containerized interface to the hardware virtualization support in Linux.
However, we found that Kata's support for SEV was too preliminary to get consistent results.
Kata or other virtualized container interfaces may provide an even simpler programming interface to SEV in the future, but they will likely have the same performance characteristics as QEMU since they both use hardware support for virtualization.
When running with QEMU, we assign all of the host cores to the guest and allocate enough memory on the guest to fit the entire resident memory of the application.
Moreover, we do not include the guest boot-up/initialization time when calculating the workload slowdown.

The documentation and scripts required to set-up and run the experiments discussed in this work are available publicly.\footnote{https://github.com/lbnl-cybersecurity/tee-hpc}

\section{Understanding the Performance of TEEs}
\label{sec:findings}
In this section, we will present our findings on the performance impact of TEEs for scientific computing workloads and the reasons \highlight{for} the observed slowdowns.
We make following main findings:
\begin{enumerate}
  \item When the user configures the NUMA allocation policy correctly, SEV has small overhead for most workloads.
  \item SEV relies on QEMU and hardware virtualization, which causes significant performance degradation for some irregular workloads, I/O intensive workloads, and workloads with high thread contention.
  \item SEV initialization is slow and depends on the memory footprint of the application.
  \item SGX has high performance overhead mostly due to its limited secure memory capacity and partially due to parallel scalability limitations and programming challenges.
\end{enumerate}

\begin{figure*}
  \centering
  \includegraphics[scale=0.8]{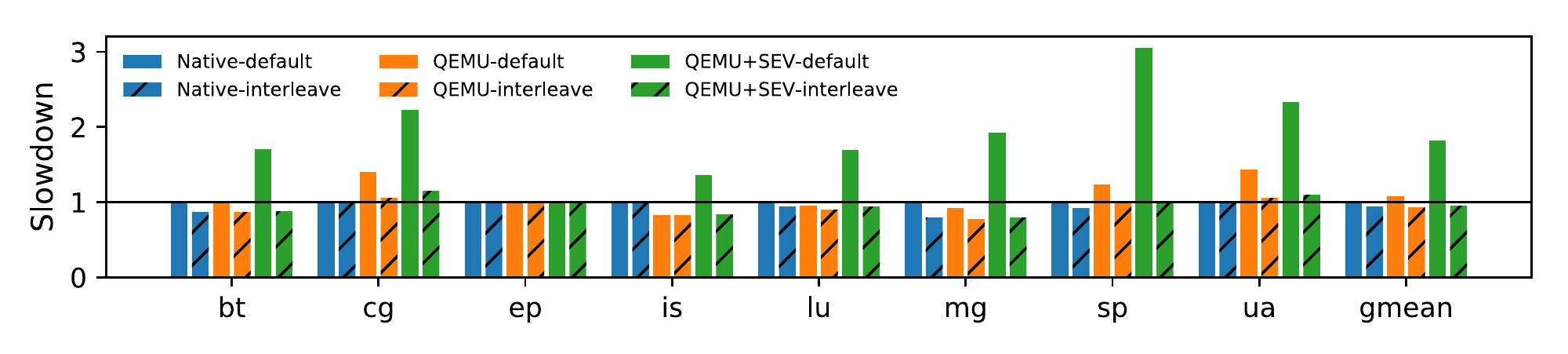}
  \caption{Performance impact of SEV for NPB C Class on AMD Naples (24 Threads). The SEV performance overhead is mainly because of default NUMA memory allocation, most of which goes away with interleaved NUMA allocation.}
  \label{fig:sevperf_npb_c}
\end{figure*}

\begin{figure*}
  \centering
  \includegraphics[scale=0.8]{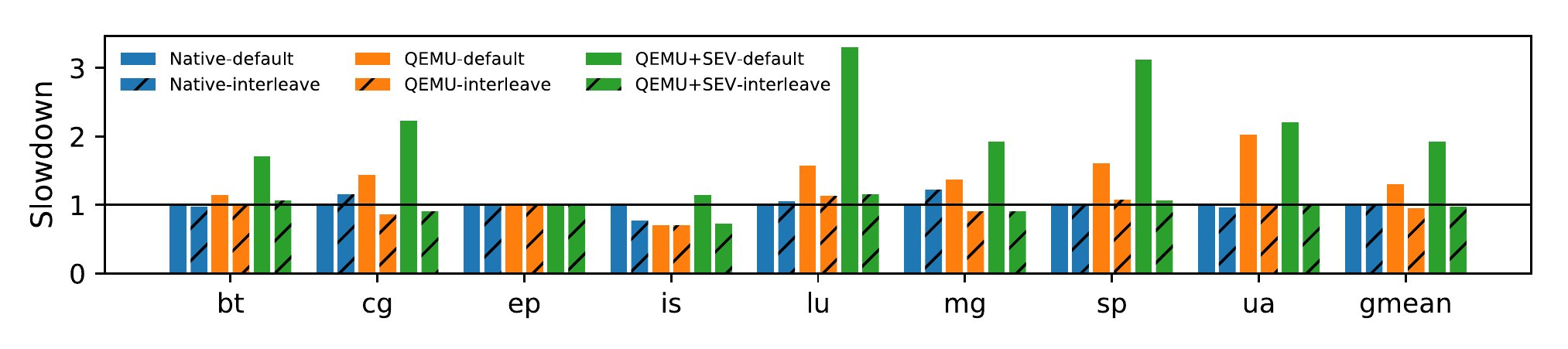}
  \caption{Performance impact of SEV for NPB D Class on AMD Naples (24 Threads).}
  \label{fig:sevperf_npb_d}
\end{figure*}

\subsection{\textbf{Finding 1:} SEV can be used for secure scientific computing without significant performance degradation for most workloads if it is configured correctly.}
\label{sec:sevperf}


SEV requires nested page tables and is only available when running in a VM.
Therefore, we compare three different cases: native (unsecure), QEMU (virtualized, but also no security gaurantees), and QEMU+SEV which provides security from the hypervisor and other users.\footnote{The initial implementation of SEV has many security vulnerabilities~\cite{du2017secure,wilke2020sevurity,li2019exploiting,werner2019severest,morbitzer2018severed,hetzelt2017security}.
However, more recent implementations (e.g., Rome) fix many of the published vulnerabilities but still have similar performance characteristics to the systems we evaluate.}

Figures~\ref{fig:sevperf_npb_c} and~\ref{fig:sevperf_npb_d} show the performance of the NAS Parallel Benchmarks for the C and D class inputs relative to the ``native'' execution without any security guarantees.
The solid bars on these figures show the performance of native execution, ``QEMU'' which is a KVM-based hypervisor running a virtual machine with the benchmark, and ``QEMU+SEV'' which has the SEV security extensions enabled (all relative to the performance of native execution).
This shows that while the performance overheads of SEV (shown in green solid bars) are lower than SGX, using the default system configuration of SEV still results in significant performance degradation compared to the virtualized QEMU execution.

In this section, we will discuss how most of these slowdowns can be eliminated through careful NUMA data placement.
We also present data from two different generations of AMD platforms to further investigate the overheads of SEV.


\begin{figure}
  \centering
  \includegraphics[width=1.0\linewidth]{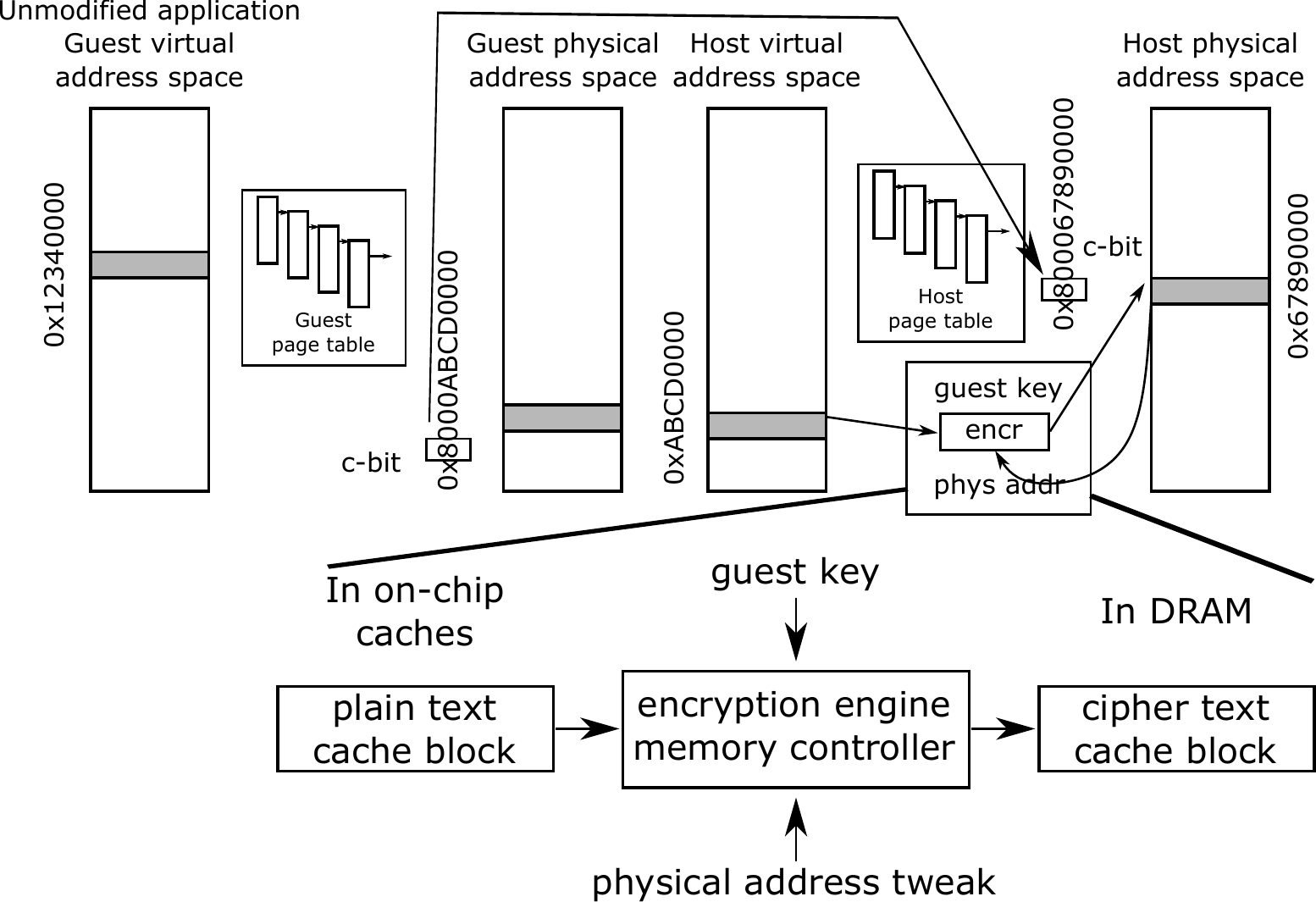}
  \caption{Details of SEV encryption implementation.}
  \label{fig:sev}
\end{figure}

\begin{figure}
  \centering
  \subfloat[SEV Default Allocation]{
  \includegraphics[scale=0.75]{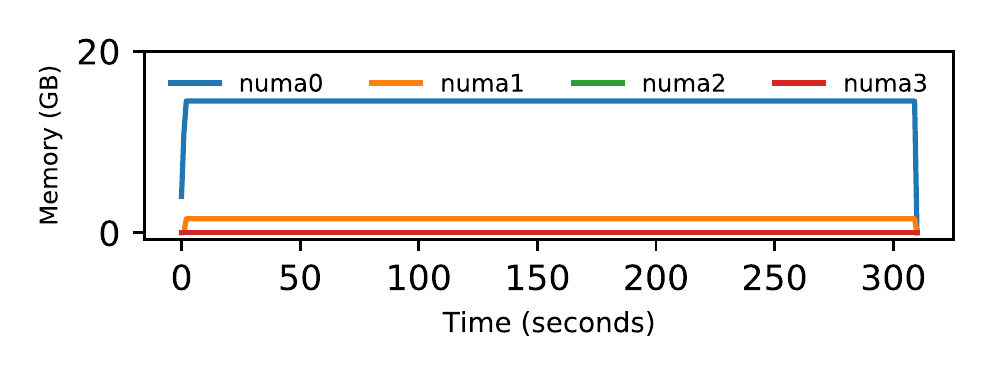}
  \label{fig:sev_def}
  }
  \newline
  \subfloat[No SEV Default Allocation]{
  \includegraphics[scale=0.75]{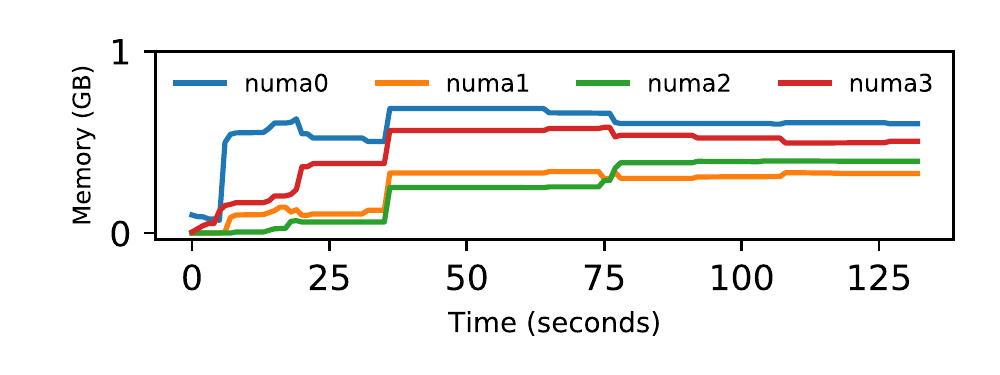}
  \label{fig:no_sev_def}
  }
  \caption{Memory allocation over time using default policy.}
\end{figure}

\textit{\textbf{Finding 1.1:} Enabling SEV causes performance degradation beyond virtualization overheads.}

Although there is some overhead from virtualization for the NAS Parallel Benchmarks as shown in the orange bars of Figures~\ref{fig:sevperf_npb_c} and~\ref{fig:sevperf_npb_d}, there is significantly more performance overhead when enabling SEV (green bars, up to 3$\times$ slowdown over the native execution).


\textit{\textbf{Finding 1.2:} SEV overhead is because of NUMA placement.}

The reason QEMU+SEV suffers more performance overhead than QEMU is that when an SEV enabled virtual machine (VM) is launched, the memory pages allocated to the guest RAM are pinned by the hypervisor (QEMU) using \emph{mlock} syscall.
As a result, all data for the application is allocated on a single NUMA node and multi-threaded processes which expect performance improvements from running on large NUMA systems suffer from performance degradation under SEV.
QEMU without SEV does not have this restriction.

\subsubsection*{Why SEV requires locking pages to physical addresses?}

Figure~\ref{fig:sev} shows details of how SEV is implemented.
This figure shows both the interaction with the nested page table translation used for hardware virtualization acceleration and the memory encryption engine.
First, this figure shows how the guest virtual address is translated through a nested page table since it must translate first into the guest physical address space then into the host physical address space.
Importantly, the ``c-bit'' or encrypted bit is removed from the guest physical address by hardware and replaced after the host page table translates the address to the host physical address space.
By removing and replacing the c-bit, the hypervisor is unaware of which pages are encrypted or not.

Second, SEV must guarantee that two identical plaintext pages present at different physical addresses in the memory will have different cipher texts to protect against cipher text block move attacks.
To make this possible, SEV uses a physical-address based tweak algorithm~\cite{du2017secure,wilke2020sevurity} as shown in Figure~\ref{fig:sev} with the physical address of the cache block influencing the cipher text via an xor-encrypt-xor tweak~\cite{rogaway2004xex}.
Since the host-physical address is used to determine the cipher-text of a page, the hypervisor cannot move a page between two physical addresses once it is allocated to the secure VM.

This limitation causes two performance issues when using SEV.
First, all data pages for the guest are \emph{pinned} in physical memory by the hypervisor~\cite{kaplan2017sev}.
In fact, because the default NUMA policy on Linux is ``first-touch'', all memory is allocated on a single NUMA node, which causes performance degradation for many of the scalable workloads evaluated in this work.
Second, SEV-based guests can under-utilize the memory resources since they do not use on-demand paging.

Figures~\ref{fig:sev_def} and~\ref{fig:no_sev_def} visualize the memory allocation process when using QEMU and QEMU+SEV.
These figures show the memory allocation over time on different NUMA nodes on a system with four NUMA nodes when a VM with 16 GB memory is launched to run (for example) \textit{sp} benchmark.
Figure~\ref{fig:sev_def} shows that under SEV all data is allocated at the time of the VM launch \emph{on a single NUMA node} as opposed to the non-SEV case (Figure~\ref{fig:no_sev_def}) which follows on-demand paging scheme and spreads the data across all four nodes.



\begin{figure}
  \centering
  \includegraphics[scale=0.65]{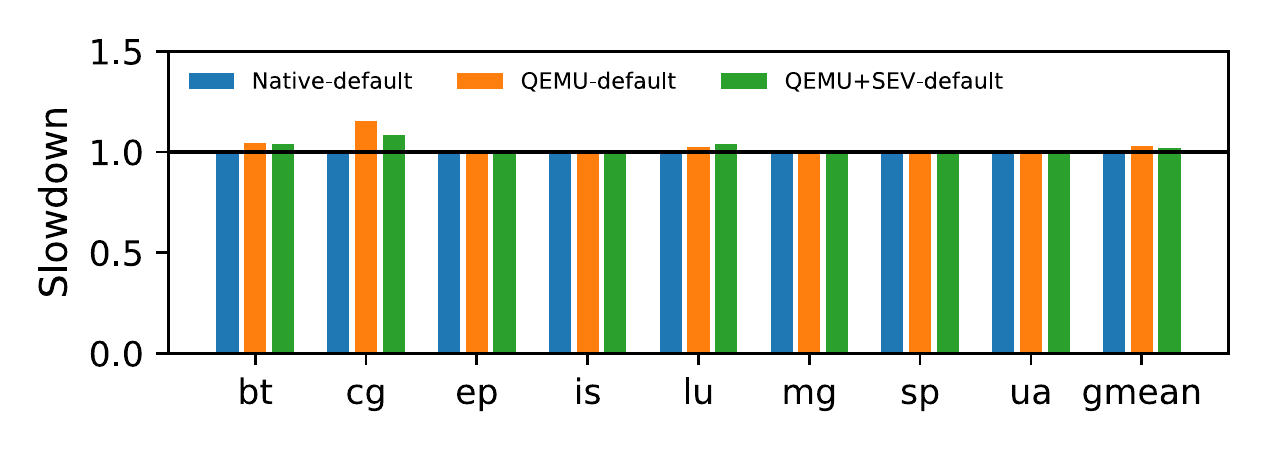}
 \caption{\highlight{NPB D Class on AMD EPYC 7402P (24 Threads)}}
  \label{fig:rome_packet}
\end{figure}

\iiswc{For additional evidence, we conducted an initial study on a single-socket AMD Rome based system (AMD EPYC 7402P, 24 core system, similar to Figure~\ref{fig:rome} but with a single package and four core dies) using NPB D class workloads.
This system has a uniform memory architecture, and that is why the slowdowns due to NUMA placement issues (observed previously) do not exist in this case as shown in Figure~\ref{fig:rome_packet}}.

Thus, we conclude the SEV-specific overhead is due to the NUMA allocation policy.

\begin{figure}
  \subfloat[SEV Interleaved Allocation]{
\includegraphics[scale=0.75]{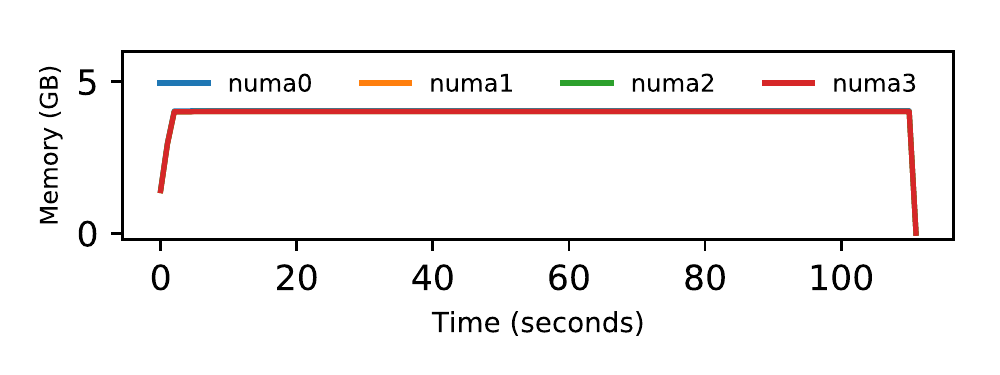}
\label{fig:sev_int}
}
\newline
\subfloat[No SEV Interleaved Allocation]{
\includegraphics[scale=0.75]{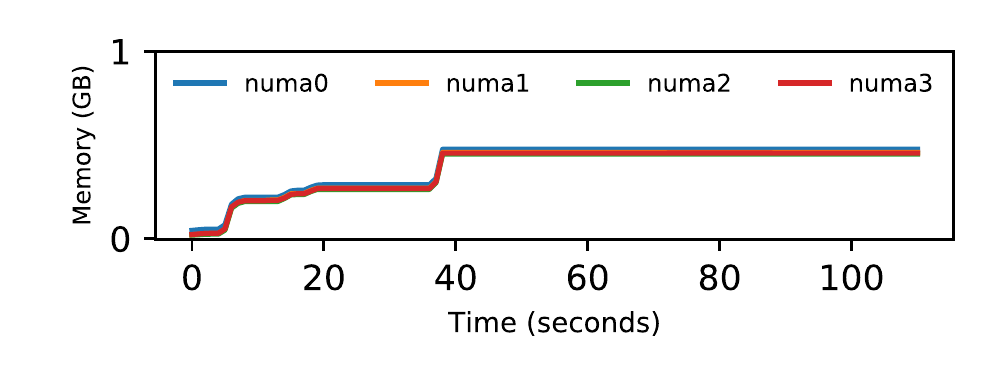}
\label{fig:no_sev_int}
}
\caption{Memory allocation over time using an interleave policy.}
\label{fig:numa_alloc}
\end{figure}

\textit{\textbf{Finding 1.3:} Explicit interleaving of data across NUMA nodes using \emph{numactl} recovers most of the performance loss.}

To mitigate the observed slowdown, we explicitly allocate memory pages across NUMA nodes rather than using the default NUMA memory allocation policy in the Linux kernel.
We use \verb|numactl| to allocate memory pages across NUMA nodes when the VM is launched under SEV.
A visualization of the memory allocation using \emph{interleaved} NUMA allocation policy is shown in Figure~\ref{fig:sev_int}.
Under SEV, an equal amount of memory (4~GB on each node) is allocated across all nodes.

We observe that the interleaved memory allocation across all NUMA nodes results in significant performance improvements for SEV.
In fact, the performance differences between QEMU and QEMU+SEV shrink as shown in Figure~\ref{fig:sevperf_npb_c} and~\ref{fig:sevperf_npb_d} when enabling NUMA interleaving (hatched bars).
This is in contrast to prior work which evaluated server-based applications and found that using a single NUMA node results in the best performance for virtualized workloads~\cite{kamio2017numa}.
\iiswc{Importantly, we also observe that for native execution the interleaved allocation results in better performance compared to the default allocation for most of the cases (prominent examples are \textit{Kripke, Mobiliti,} and \textit{cc} from GAPBS)}

\begin{figure*}
  \centering
  \subfloat[GAPBS (road network)]{
  \includegraphics[width=.62\textwidth]{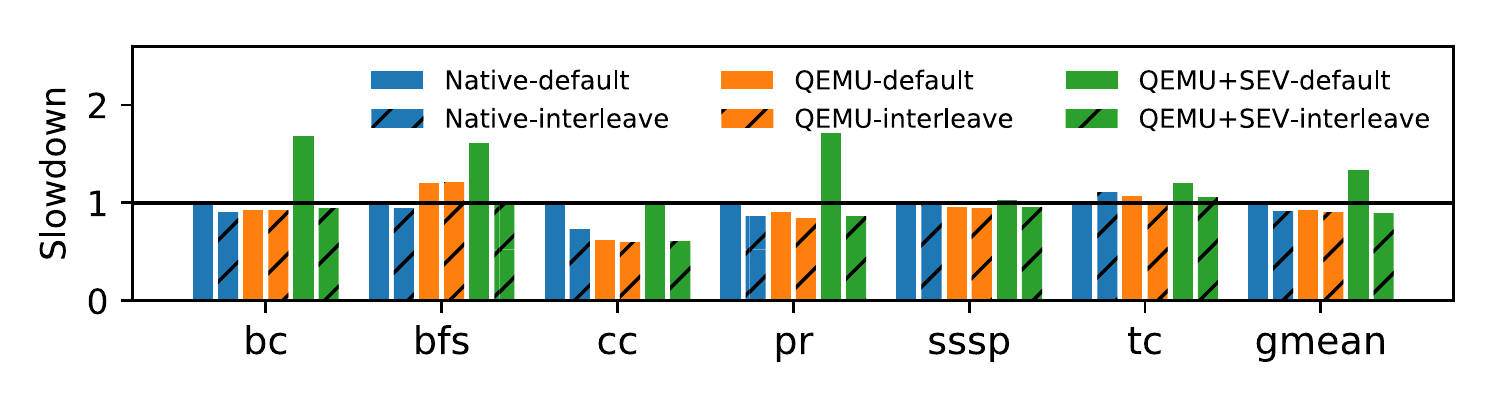}
  \label{fig:gapbs1}
  }
  \subfloat[Real world HPC workloads]{
  \includegraphics[scale=0.75]{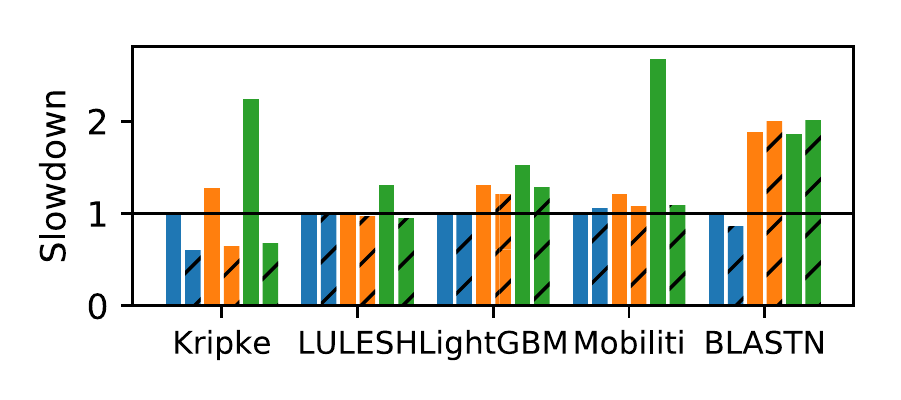}
   \label{fig:realhpc}
  }
  \caption{Performance impact of SEV for GAPBS and other real world HPC workloads on AMD Naples (24 Threads). Interleaved NUMA allocation works for graph and other HPC workloads as well except BLASTN which shows high overhead mainly because of virtualized disk I/O operations.
}
\end{figure*}

In addition to the HPC kernels in the NAS Parallel Benchmarks, we also studied modern HPC workloads.
Figure~\ref{fig:gapbs1} shows the execution time for native, QEMU and QEMU+SEV cases for GAPBS workloads when executed using a road network graph.
Similar to NPB, NUMA interleaving reduces the difference between QEMU+SEV and SEV.
Similar trends are found for other HPC workloads as shown in Figure~\ref{fig:realhpc}.

However, there are still some cases where QEMU and QEMU+SEV experience performance degradation compared to the native (unsecure) baseline.
These differences can be attributed to \textit{virtualization} overhead as discussed in Finding~2.


\begin{figure*}
\centering
\includegraphics[scale=0.9]{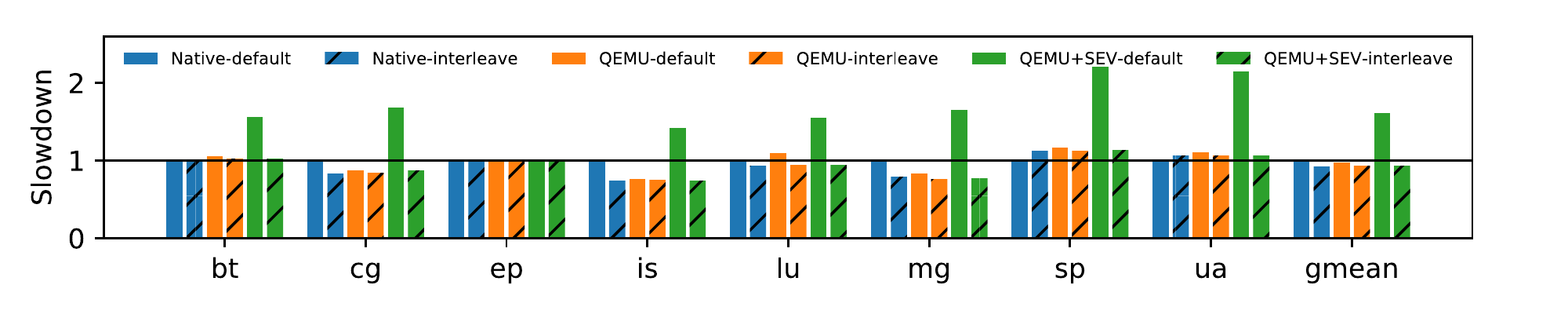}
\caption{Performance impact of SEV for NPB D Class on AMD Rome (128 Threads)}
\label{fig:sevperf_npb_d_shasta}
\end{figure*}

\begin{figure*}
  \centering
  \subfloat[GAPBS (road network)]{
  \includegraphics[width=.61\textwidth]{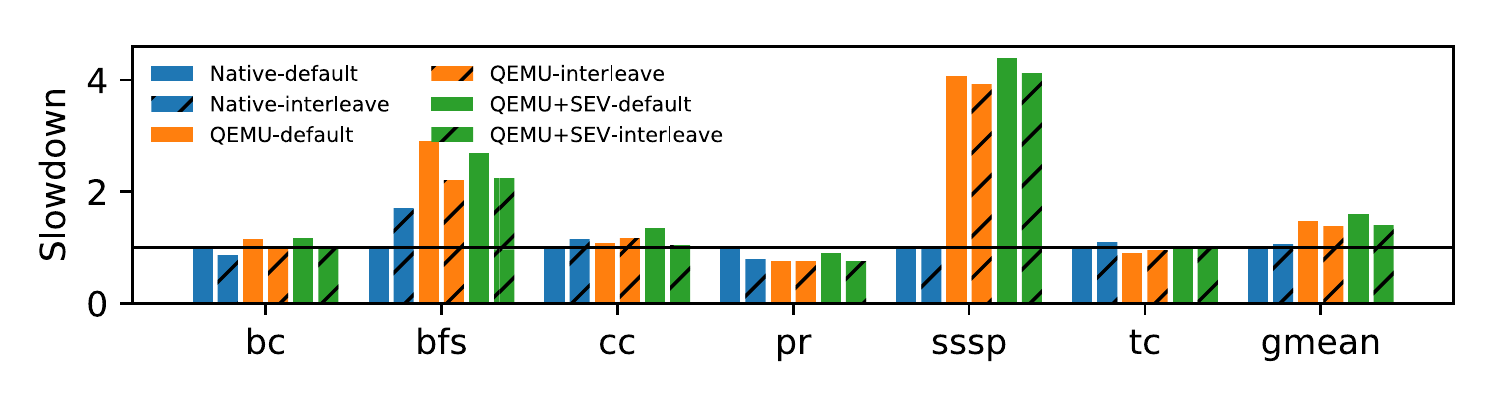}
  \label{fig:gapbsroad_shasta}
  }
  \subfloat[Real world HPC workloads]{
  \includegraphics[scale=0.75]{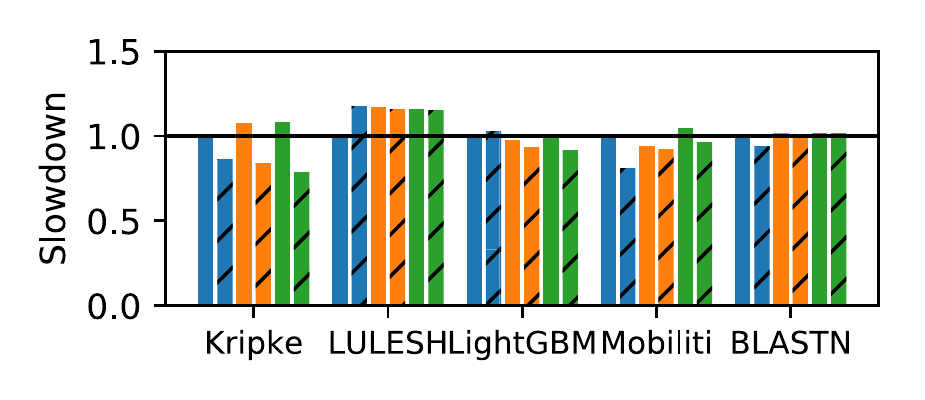}
   \label{fig:realhpc_shasta}
  }
  \caption{Performance impact of SEV for GAPBS and other real world HPC workloads on AMD Rome (128 Threads). NUMA placement still matters on platforms with more uniform memory architecture. Two examples where main cause of overhead is virtualization are bfs and sssp.}
\end{figure*}

\textit{\textbf{Finding 1.4:} NUMA placement still matters on new platforms with more uniform memory architecture.}

As discussed in Section~\ref{methodology}, we studied the performance of these benchmarks on another modern server class AMD machine EPYC 7702 (Rome architecture), which contains 2 NUMA nodes instead of four (see Figure~\ref{fig:numa}).
Figure~\ref{fig:sevperf_npb_d_shasta}, Figure~\ref{fig:gapbsroad_shasta} and Figure~\ref{fig:realhpc_shasta} show the relative performance of native, QEMU, and QEMU+SEV for the Rome system.
Similar to the Naples system, there are significant overheads when using SEV unless the data is explicitly interleaved between NUMA nodes.
Thus, even for systems that have more ``uniformity'' in their memory architecture, data placement is important for performance when using SEV.

\textit{\textbf{Finding 1 summary:}}
When enabling SEV, there are additional overheads beyond just the virtualization platform overheads.
These overheads are caused by the memory allocation restrictions of the SEV technology and persist even on the most recent architectures.
However, we can overcome these SEV-specific overheads by explicitly interleaving data between NUMA nodes when the virtual machine is initialized.

\subsection{\textbf{Finding 2:} The remaining SEV performance differences are due to virtualization overheads.}
We find that in some cases there is performance degradation of the QEMU+SEV system compared to the baseline native execution even after applying our NUMA interleaving configuration change.
These slowdowns come from the use of hardware virtualization and QEMU.
For example, in Figure~\ref{fig:gapbsroad_shasta}, \textit{sssp} with QEMU+SEV shows considerable slowdown compared to Native-default case irrespective of memory allocation policy (default or interleaved) on AMD Rome architecture.
As visible in the Figure~\ref{fig:gapbsroad_shasta}, the performance of QEMU+SEV and QEMU match, indicating that the main cause of this slowdown is virtualization itself, not the SEV extension.

We observed that, when run with 128 threads (as in Figure~\ref{fig:gapbsroad_shasta}), \textit{sssp} shows much higher number of kvm exits per second caused by the PAUSE instruction in comparison to the case when it is run with a smaller number of threads (e.g., 32).
The PAUSE instruction is used to implement spinlocks and can cause KVM exits (i.e., a usermode to hypervisor switch) which has a higher latency than a normal context switch.

In fact, when executed with only 32 threads, the virtualization slowdown of \textit{sssp} improves to 1.7$\times$ (in contrast to 4$\times$ in Figure~\ref{fig:gapbsroad_shasta}).
Similarly, the QEMU overhead for \textit{bfs} reduces to 1.6$\times$ with 32 execution threads in contrast to 2.6$\times$ with 128 execution threads (Ding et al. made similar findings~\cite{ding2013hidden}).
Thus, when using QEMU or QEMU+SEV it is important to use the appropriate number of execution threads for your workload and workloads with highly contended locks may result in significant performance degradation.

In Figure~\ref{fig:realhpc}, BLASTN also shows slowdown by virtualization on AMD Naples architecture.
The nucleotide database which is used by BLASTN is approximately 245GB in size (much larger than the memory size of 64~GB on our AMD Naples system), which leads to many disk I/O operations and thus slowdown under virtualization.
\highlight{On the other hand, when the same workload is executed on AMD Rome system (which has 1~TB of memory), there is not any noticeable virtualization overhead as shown in Figure~\ref{fig:realhpc_shasta} since the workload can fit in the available system memory.}

There is significant prior work quantifying the impact of virtualization on the performance of HPC workloads~\cite{ding2013hidden,jackson2010performance,younge2011analysis,ha2016impact,gupta2013and,kudryavtsev2012modern,gandhi2014efficient}.
These prior works mostly focus on overheads from TLB misses and nested page table walks.
Similarly, our results show the virtualization overheads grow as the working set of the applications grow and are worse for workloads with irregular access patterns (e.g., graph workloads).
Prior work has shown you can reduce this overhead by using huge pages or through changes to the hardware (e.g., Virtualized Direct Segments~\cite{gandhi2014efficient}).
Additionally, the work of Ding et al.~\cite{ding2013hidden} presents possible strategies to mitigate the virtualization slowdown caused by multithreaded application scaling.

\subsection{\textbf{Finding 3:} SEV initialization is slow and depends on the memory footprint of the VM.
Note: all previous data ignores VM initialization time.}
We find that the time taken to initialize the workload is significant when using QEMU and increases when using QEMU+SEV.
When using QEMU or QEMU+SEV, before running the workload the virtual machine guest operating system must complete the boot process.
For QEMU this bootup time takes about one minute for our workloads.

\begin{figure}
  \centering
  \includegraphics[scale=0.7]{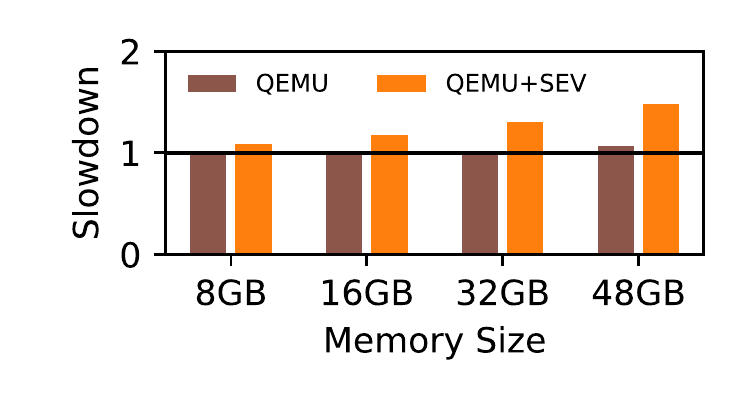}
  \caption{Performance of VM boot (relative to QEMU-8GB)}
  \label{fig:vmboottime}
\end{figure}

However, when enabling QEMU+SEV, this boot time increases due to the hypervisor having to initialize the memory before handing it over to the guest OS.
As shown in Figure~\ref{fig:vmboottime}, SEV can cause a slowdown (relative to QEMU-8~GB)
of 1.1$\times$--1.47$\times$ depending on the size of the VM memory (from 8~GB to 48~GB).
In addition to the memory initialization, QEMU+SEV also needs extra time for key management when launching a guest with QEMU+SEV.
However, we believe that the main source of SEV slowdown is the fact that the entire VM memory has to be allocated at once in case of QEMU+SEV in contrast to on-demand allocation in case of QEMU (as discussed in section~\ref{sec:sevperf}), as evident by the increase in slowdown as the VM memory size is increased.
This can specially become a bottleneck for the use cases where the user intend to launch their jobs in a new VM each time (e.g., when using Kata containers~\cite{kata}).

\begin{figure*}
  \centering
  \subfloat[Relative performance running under SGX compared to native execution.]{
  \includegraphics[width=.5\textwidth]{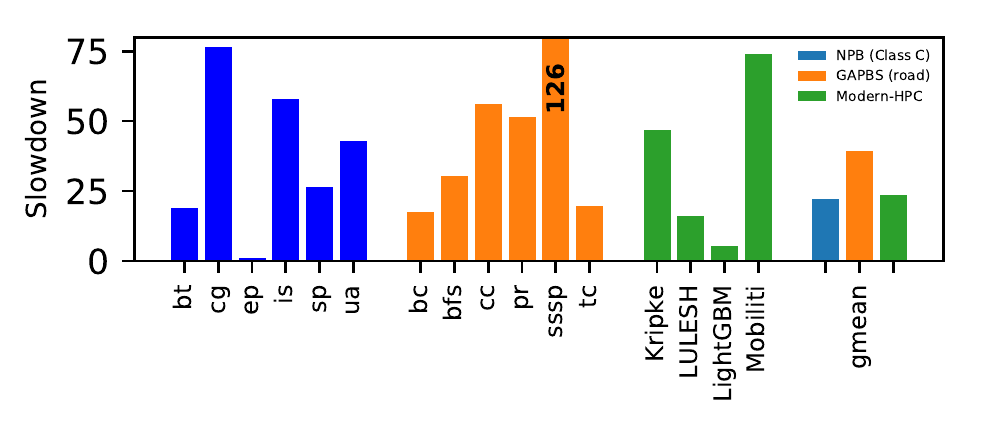}
  \label{fig:sgxperf}
  }
  \subfloat[EPC Fault Rate (Per Million Instructions) when running under SGX]{
  \includegraphics[width=.5\textwidth]{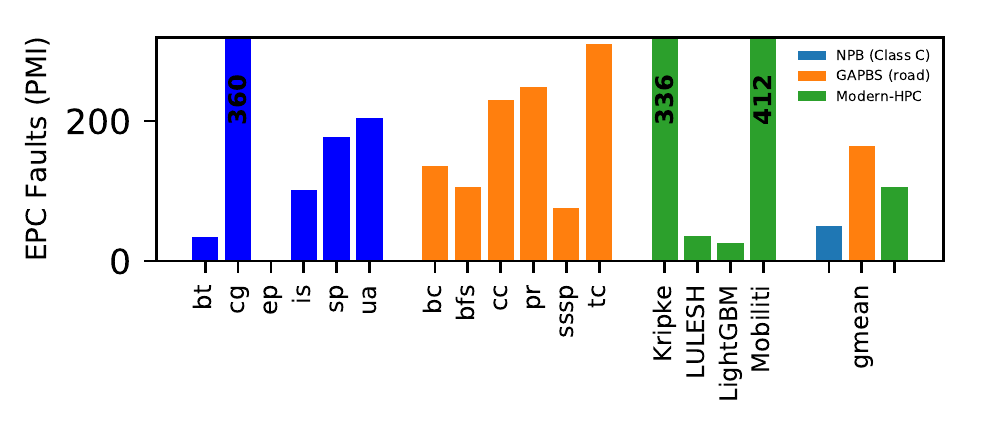}
  \label{fig:sgxepc}
  }

  \caption{Performance Impact of SGX and its Relation to EPC Faults}
  \end{figure*}

\subsection{\textbf{Finding 4:} SGX is inappropriate for \highlight{unmodified} scientfic computing applications.}
\label{sec:sgxperf}

We find a number of reasons that SGX is not an appropriate technology for securing HPC workloads.
A primary design goal of SGX is to enable a small trusted compute base, and SGX was not designed to support large scale workloads.
We find that running HPC workloads under SGX causes a (1$\times$--126$\times$) slowdown (mostly due to its limited secure memory capacity), workloads exhibit poor thread scalability under SGX, and it is difficult to adapt HPC code to work under the SGX programming model.

\textit{\textbf{Finding 4.1:} Workloads with working sets larger than about 100~MB suffer large performance degradation under SGX.}
Figure~\ref{fig:sgxperf} shows the slowdown of HPC workloads under SGX compared to an un-secure baseline.
For this experiment, we ran NPB with the ``class C'' inputs (blue in Figure~\ref{fig:sgxperf}).
We were limited to using the class C inputs, as most class D inputs were too large to run on the desktop systems that support SGX.
However, we believe that running larger inputs under SGX would show at least as much performance overhead as the smaller inputs.
We also show the relative performance of graph workloads and other modern HPC workloads in Figure~\ref{fig:sgxperf}).
We were not able to run BLASTN workload with SGX due its dependencies (discussed more in Finding~4.4).

Most of the performance degradation shown in Figure~\ref{fig:sgxperf} can be explained by the overhead of moving data from un-secure memory into secure memory.
SGX has a limited amount of secure memory, about 100~MB.
Thus, any workload with a working set larger than 100~MB must use the secure memory as an \emph{enclave page cache} (EPC).
The EPC is managed by the SGX driver in software and has similar behavior to OS swapping and moving pages between normal and secure memory is a high latency event.

Figure~\ref{fig:sgxepc} shows the number of EPC faults per million instructions for each of the workloads.
This figure shows that most of the slowdown in Figure~\ref{fig:sgxperf} can be explained by the EPC fault rate.
The workloads with the highest rate of moving data between secure memory and normal memory (e.g., \textit{cg} from NPB, \iiswc{Mobiliti, and Kripke) show very high slowdown}.
On the other hand, \textit{ep} from NPB shows little performance overhead with SGX because it has a very small working set size (about 28~MB) which fits in the EPC and does not require data movement between secure and normal memory spaces.

\textit{\textbf{Finding 4.2:} In some cases, SGX slowdown can be caused by system calls.}
Applications under SGX exit the enclave to process a system call.
This can become a problem for workloads with a large number of system calls.
In the studied HPC workloads, the only case where we found system calls to be the dominant source of performance overhead is \textit{sssp} benchmark from GAPBS.
As shown in Figure~\ref{fig:sgxepc}, the slowdown for \textit{sssp} does not correlate with EPC fault rate.
\textit{sssp} shows significantly higher number of enclave exits (and system calls) and is the main contributor to its performance overhead compared to the un-secure execution.
\highlight{Most of these system calls were \textit{write} and \textit{futex} calls, which are needed due to the benchmark printing progress to the terminal.
The \textit{futex} calls are used for synchronization (of multiple threads) before printing the status messages using \textit{write} calls.
The effect on slowdown because of \textit{futex} system calls can be understood by the difference in the observed slowdown for six thread execution (126$\times$) and single thread execution (20$\times$), which does not need any synchronization.
}

\begin{figure}
  \centering
  \subfloat[cg]{
  \includegraphics[width=.24\textwidth]{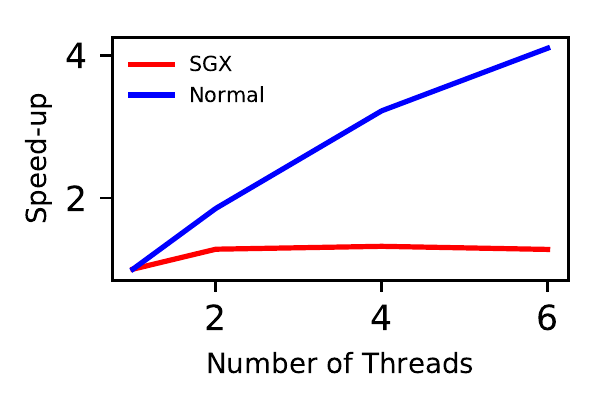}
  \label{fig:threadscg}
  }
  \subfloat[ep]{
  \includegraphics[width=.24\textwidth]{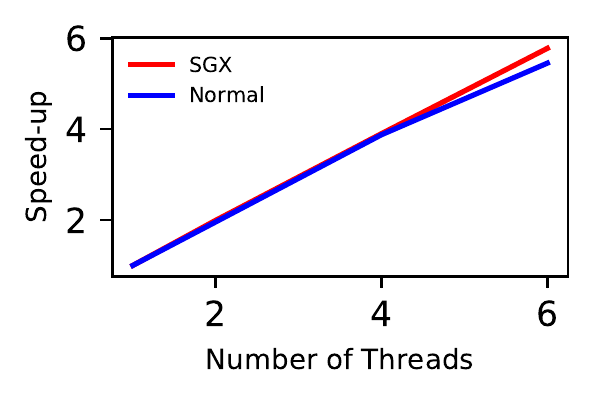}
  \label{fig:threadsep}
  }
  \caption{Impact of Multiple Execution Threads}
  \label{fig:threads}
\end{figure}

\textit{\textbf{Finding 4.3:} Workloads exhibit poor multithreaded scaling under SGX.}
Another factor that aggravates the slowdowns under SGX is explained with the help of Figure~\ref{fig:threads}, which shows the workload performance when increasing the number of threads.
Figure~\ref{fig:threadscg} shows that \textit{cg} only achieves a speedup of 1.4$\times$ with six threads when using SGX compared to about 4$\times$ speedup normally.
We hypothesize that the handling of EPC faults by the SGX kernel driver becomes the serializing factor because all logical processors executing an enclave's code are required to exit the enclave whenever an EPC page is deallocated~\cite{costan2016intel}.
Similar behavior is exhibited by most of the other workloads with high resident memory size.
On the other hand, workloads with working set sizes that fit in the EPC (e.g., \emph{ep}) scale under SGX as they would under normal execution as shown in Figure~\ref{fig:threadsep}.

\textit{\textbf{Finding 4.4:} SGX's programming model is a poor fit for HPC applications.}
Intel distributes an official SDK~\cite{sgx-sdk} for SGX which requires users to re-write their application and divide it into two pieces, secure code and non-secure code.
Due to the complex nature of HPC codes, dependencies on external libraries, and frequent use of legacy codes (including a non-trivial number of them written in Fortran), we investigated several alternative interfaces to SGX which reduce the burden on the programmer.

There are multiple third party solutions to run unmodified applications under SGX including SCONE~\cite{arnautov2016scone}, Graphene~\cite{Tsai2017-sp}, Haven~\cite{baumann2015shielding} and Asylo~\cite{asylo}.
We did initial experiments with both Graphene and SCONE as they were the best supported third party solutions at the time we ran our experiments.
SCONE provides containerized environment and is easier to set-up and has a better support of running diverse workloads without any modifications, so we used SCONE for our experiments.
Although we only evaluated SCONE, all SGX programming interfaces have similar limitations due to SGX's design which limits the TCB.
Graphene~\cite{Tsai2017-sp} is not as convenient to use as SCONE and Google Asylo's~\cite{asylo} recently added support to run unmodified applications still lags behind SCONE in terms of the number of supported use-cases.
Open Enclave SDK~\cite{oenclave} is another SDK to build enclave applications and does not support unmodified applications.

We found that even with SCONE, which promises to run unmodified applications with SGX, it is fundamentally difficult to use SGX to run HPC applications.
In order to keep the library OS simple, SCONE makes use of the musl libc library,
instead of more traditional C library glibc, along-with some containerized services using the
Linux Kernel Library (LKL).
The use of musl libc instead of glibc means many applications are not portable to SCONE (e.g., BLASTN failed to compile inside SCONE  and many common frameworks such as TensorFlow require glibc instead of musl libc).
Moreover, SCONE does not support some system calls like fork, exec and clone mainly due to its user space threading model and the architectural limitations of SGX~\cite{arnautov2016scone} which further limits its applicability to scalable applications.

\textit{\textbf{Finding 4 summary:}}
The current implementation of Intel's SGX limits the secure memory size which severely affects the performance of any workload that has a working set that does not fit in this cache.
Additionally, there is currently no stable support to run unmodified workloads under SGX.
The limited EPC capacity and application partitioning are a fundamental design constraints of SGX.
Thus, we conclude that SGX is unsuitable for secure execution of HPC applications.
In Section~\ref{sec:conclusion}, we discuss possible future SGX implementations and how they may improve performance for some HPC workloads.

\section{Discussion}
\label{sec:discussion}

\subsection{Beyond Single Node}
Scientific computing workloads often scale across multiple machines (nodes).
In this work, we only focus on a single node to isolate the performance impact of hardware TEEs.
To understand the impact of communicating between TEEs on multiple nodes, we conducted a preliminary investigation of a multi-node system with support of SEV on CloudLab~\cite{Duplyakin+:ATC19}.

Current HPC systems mostly rely on high performance transport protocols like RDMA for communication among multiple nodes.
However, RDMA does not provide any secure communication support although there is a recent research proposal for secure RDMA~\cite{taranov2020srdma}.
Therefore, we instead evaluated TCP for communication among machines using OSU MPI microbenchmarks~\cite{osu-mpi}.
For point-to-point bandwidth benchmarks, we observed a 2$\times$ reduction in bandwidth  when comparing QEMU to QEMU+SEV, but
the latency remains the same (approximately 1000 $\mu$s) for both QEMU and QEMU+SEV (ranging from 1 byte to 2MB packet sizes).

However, in this simple benchmark, the communication between nodes is insecure as there is no support for encrypted communication between multiple nodes in SEV automatically.
A naive solution to make this communication secure is to use a VPN.
We experimented with OpenVPN and found the slowdown of VPN based secure communication to be large.
For example, for above microbenchmarks, the bandwidth number drops over 10$\times$ and
latency increases by almost $20\times$.

Therefore, we conclude that there is a need to develop more performant architectures to enable TEEs across multiple nodes in a distributed memory.
There are some existing software based solutions to enable encrypted communication across nodes, but they might not be sufficient for scientific computing scale workloads.
For example, SCONE~\cite{arnautov2016scone} provides a network shield which transparently protects the communication among multiple nodes (each with its SGX hardware).
Asylo~\cite{asylo} (open-source framework for developing enclave applications) allows enclave based applications to be scaled across multiple machines using gRPCs (google remote procedure calls), while being agnostic to TEE implementation.

\subsection{TEEs in Accelerator Based Systems}
As computer systems become more heterogeneous and employ different hardware accelerators to optimize for specific applications, the attack vector expand in size.
Thus, it is important to expand TEEs across accelerators and main processing elements in such systems.
Research on TEEs in accelerator based systems is still at very early stage.
Examples of research works to enable trusted execution in accelerator based systems include:
Graviton~\cite{volos2018graviton} and HETEE~\cite{hetee}.
Graviton~\cite{volos2018graviton} provides trusted execution capabilities on GPUs by providing isolation to sensitive kernels from other code executing on GPUs and the host.
HETEE~\cite{hetee} relies on a centralized controller to manage trusted execution on all computing units including GPUs and accelerators.
However, there is still a need to investigate what new features need to be added to existing practical TEE solutions to enable trusted execution in accelerator based systems.

\subsection{Observations on Security of TEEs}
Although the focus of this paper is the performance analysis of TEEs for secure HPC, we provide a brief security analysis of TEEs discussed in this paper.
SGX provides integrity guarantees while SEV lacks such support.
However, the weaker guarantees of SEV are considered to be good enough by Google's confidential cloud computing initiative, and these guarantees are becoming stronger.
AMD has introduced  SEV-ES~\cite{amd:ref} that adds encryption of guest register state
to provide additional protection against VM state related attacks, and SEV-SNP~\cite{amd:ref}
 provides integrity checks.
These are encouraging developments from the security perspective,
as they address some of the vulnerabilities and limitations of SEV.
It should be noted that SEV-SNP~\cite{amd:ref} does not provide integrity guarantees using Merkle tree like data structures (as SGX does).
Therefore, it is more scalable and can support larger secure memory sizes.
Additionally, it seems Intel is also moving in the direction of full memory
encryption and virtual machine-based trusted execution environments like AMD's SEV with total memory encryption (TME) and multi-key total memory encryption (MKTME) technologies~\cite{intel:memencrypt,intel:tme}.

Finally, in this paper we have focused on just one aspect of the entire secure application workflow which may include other steps as well (like a secure connection to the computing resources) in addition to running it inside an SEV or SGX enclave.
However, we believe that the execution of the workload itself in a secure enclave is the most important factor for performance analysis.

\subsection{Research Avenues}
Next, we discuss few research avenues in the context of the use of TEEs to enable  secure HPC.
\newline
\subsubsection{Research Avenues for Software Frameworks}
First we discuss some research directions for software frameworks:
\newline
\newline
\noindent
\textbf{\textit{Intelligent job scheduling:}} Security requirements of HPC workloads can be diverse.
At the same time, the HPC platforms can be heterogeneous possibly composed of nodes from multiple vendors e.g., Intel and AMD, thus making both SGX and SEV available in the same environment.
Therefore, we propose the idea of an intelligent job scheduler which can allocate applications to an appropriate node depending on their sensitivity and the expected slowdown from the secure environment.
The sensitivity of the workload can be fed to the scheduler from the user and the expected slowdown can be calculated using pre-trained models.
As a proof of concept, we train a second-degree regression model to predict the slowdown of workloads under SGX, using only the features from normal execution of programs.
The slowdown under SEV can be assumed to be negligible if interleaved NUMA allocation is used as we showed in our performance analysis.
For SGX, we can achieve a mean square error of \num{2.81e-11} (an R2 score of 0.99) for a second-degree regression model.
The features (normalized to per million instructions) used in the model include: resident\_memory, native\_time, syscalls, loads, and mem\_accesses in the order of the highest to the lowest observed correlation (with the slowdown).
For SGX, the highest correlation with slowdown is shown by resident\_memory feature which is directly related to EPC faults which appears to be the primary reason for SGX slowdowns (section~\ref{sec:sgxperf}).
\newline
\newline
\noindent
\textbf{\textit{Automatic Application Partioning:}}
Partitioning of applications into secure and non-secure parts is a difficult task, especially in HPC settings as HPC workloads often rely on various third-party libraries.
Thus, there is a need for support of automated porting of applications to secure environments and their partitioning into secure and non-secure parts.
For instance, Lind et al. \cite{Lind2017-un} have already proposed Glamdring which is a framework for automatic application partitioning into secure and non-secure parts.
Automatic application partitioning not only makes it easier to use secure environments, it can also help in mitigating the performance slowdowns by keeping the secure memory footprint to the minimum.
\newline
\newline
\noindent
\textbf{\textit{Dynamic Migration of Application Segments:}}
Another research direction to explore is building of tools to shift sensitive functions or parts of the application transparently to enclaves at runtime.
Similarly, the outsourcing of unsecure parts of the application to (untrusted) accelerators or general cores to improve performance can be done transparently using tools (if developed).
\newline
\newline
\noindent
\textbf{\textit{KVM exit clustering for mitigating slowdown:}} For virtual machine based TEEs, like AMD SEV, by scanning and decoding upcoming instructions the hypervisor can identify the ones that will cause a KVM exit.
Then a cluster of exiting instructions can be formed which can be executed all at once, ultimately reducing the overall exit rate and the cycles spend for saving/restoring the VM's state leading to improved performance.
\newline

\subsubsection{Research Avenues for Computer Architects}
Following are some research possibilities to explore for hardware developers to enable more optimized trusted execution in HPC settings:
\newline
\newline
\noindent
\textbf{\textit{Secure Memory Size:}}
  The limited secure memory size (as in SGX) is an issue that can have a severe impact on the
  performance of trusted execution of HPC workloads.
  One of the biggest hurdles in increasing the size of the EPC (secure memory) in SGX is the cost associated with the
  metadata that is needed to provide security guarantees of confidentiality and integrity.
  SGX maintains an integrity tree, composed of large counters, to keep track of the version of EPC pages to protect against replay attacks.
  We believe that there is an opportunity to optimize the design of integrity trees along with the other metadata to enable the bigger size of secure memory.
  For example, recent research by Taassori et al. \cite{taassori2018vault} introduced a variable arity integrity tree (VAULT) that results in a compact design with low depth.
  Saileshwar et al. \cite{Saileshwar2018-pu} proposed a compact integrity tree that uses morphable counters rather than fixed-size counters to
  accommodate more counters in the same memory area.

  Currently in Intel SGX, whenever there is an EPC page fault, it is handled by a fault handler in the SGX kernel driver.
  This handling is a very expensive process (also requires the logical control flow to exit the enclave), which leaves room for some kind of hardware-based acceleration of this fault handling.
  An example of research efforts to reduce this cost is the work of Orenbach et al. \cite{Orenbach2017-ks}, Eleos, which uses software-based address
  translation and management inside the enclave to eliminate the need of flowing out of the enclave in case of a page fault.
  Similarly, there lies an opportunity to explore the dynamic remapping of pages rather than actually copying them from one memory type to another.
\newline
\newline
\noindent
\textbf{\textit{Intelligent Page Prefetching and Eviction:}}
 By learning the memory access pattern of applications, the sensitive pages can be prefetched in the secure memory from non-secure
 memory regions even before they are actually accessed, thus reducing the number of access faults.
 Currently, before control is transferred to the kernel driver to handle EPC fault, lower 12 bits of CR2 register (holding the faulting virtual address) are cleared because of security concerns.
Thus the driver is not able to use those bits to help in predicting memory access patterns.
Moreover, Gjerdrum et al. \cite{gjerdrum2017performance} have shown the page fault handler to be over-eager and unable to utilize EPC exhaustively.
There lies an opportunity to enable kernel driver to use the application's memory access patterns to prefetch pages anticipated to be used or perform smart page eviction.
Similarly, the pre-fetching and page eviction can be handled entirely in the hardware making it part of the TCB.

\section{Conclusion}
\label{sec:conclusion}

In this paper, we studied the performance impact of two TEEs (SEV and SGX) for secure execution of scientific computing workloads.
We showed that while Intel's SGX causes significant performance degradation for \highlight{unmodified} HPC workloads, when configured correctly AMD's SEV technology can provide protection against the most important HPC threats with minor performance overhead.
Although there is no hardware available with Intel's MKTME technology, we believe the performance characteristics will be similar to the QEMU+SEV performance presented in this paper since these two technologies have closely related designs.
The new developments in AMD SEV (SEV-ES\cite{amd:ref} and SEV-SNP\cite{amd:ref})
do not change the fact that pages are still pinned to a given NUMA node, so the
issues discussed in section~\ref{sec:findings} still remain.

So, we believe, architectural innovations are necessary to mitigate the overheads reported in section~\ref{sec:findings}.
Specifically, we are planning to harness the ideas in direct segments and coalesced
TLBs~\cite{gandhi2014efficient,kunati2018implementation}
to mitigate the overhead of virtualization in SEV.
To reduce the overhead of SGX we are exploring opportunities for more intelligent page
prefetching and eviction by learning the memory access pattens of applications and implementing it
in hardware as part of the TCB.

\section*{Acknowledgement}

This work was supported by the Director, Office of Science, Office of Advanced Scientific Computing Research, of the U.S. Department of Energy under Contract No. DE-AC02-05CH11231 and the National Science Foundation under Grant No. CNS1850566.
Any opinions, findings, conclusions, or recommendations expressed in this material are those of the authors and do not necessarily reflect those of the sponsors of this work.

\bibliographystyle{IEEEtran}
\bibliography{ref}

\end{document}